\documentclass[prd,twocolumn,aps,eqsecnum]{revtex4}
\begin{document}
\title{  A comparison of  search templates
 for gravitational waves from binary inspiral --- 3.5PN update}

\author{Thibault Damour}

\affiliation{\it Institut des Hautes Etudes Scientifiques, 91440
Bures-sur-Yvette, France}

\author{Bala R. Iyer }
\affiliation{\it Raman Research Institute, Bangalore 560 080, India}

\author{B.S. Sathyaprakash}
\affiliation{\it School of Physics and Astronomy,
Cardiff University, 5, The Parade, Cardiff, CF24 3YB, U.K.}
\date{\today}

\begin{abstract}
Phasing formulas in \cite{DIS3}
 are updated taking into account the recent 3.5PN results.
 Some  misprints in the published version of
 \cite{DIS3}  are also corrected.
\end{abstract}

\maketitle

\section {Introduction}
In this note  we update Tables I and II in Ref. \cite{DIS3} (henceforth
referred to as DIS3) in view of the recent theoretical progress
made in the dynamics of, and radiation from, binary systems
to 3.5PN order \cite{DJS00,BF00,DJS01,BFIJ01}.
Before giving a comprehensive list of the corresponding updates, we
take this opportunity to correct some misprints in Ref. \cite{DIS3}. 
Results in \cite{DIS3}  are not changed since they
used the correct formulas  free of the misprints below.

The expansion coefficients 
in Table I and Table II of DIS3, are all Newton-normalised 
coefficients.  In the notation of the paper, there should 
be overhats on all these coefficients, except $e_k$ (that is,
$\widehat E_k,$ $\widehat {\cal F}_k,$ $\hat t^v_{k},$ 
$\hat \phi^v_k,$ $\hat \phi^t_k,$ $\widehat F^t_k,$ and $\hat \tau_k$).
The coefficients $\hat t^v_5,$ $\widehat F^t_5$ and $\hat \tau_2$ 
in Table II (which were correct in the eprint
version) contain typographical errors in the published version
of DIS3. Their correct expressions
are:
\begin{equation}
\hat t^v_5 = - \left ( \frac {7729}{252} - \frac{13}{3}\eta \right )\pi,
\end{equation}

\begin{equation}
\widehat F^t_5 = - \left(\frac{7729}{21504} - \frac{13}{256}\eta\right)\pi,
\end{equation}

\begin{equation}
\hat \tau_2 = \frac{5}{9} \left ( \frac{743}{84} + 11\eta\right ).
\end{equation}

The second  of the equations in Eq.~(4.5) of DIS3 should read 
\begin{equation}
p^0_\varphi = \left [\frac {r_0^2 - 3 \eta}{r_0^3 - 3 r_0^2 + 5 \eta} \right ]^{1/2} r_0, 
\end{equation}
the factor $r_0$ outside the square brackets on the 
right-hand side is missing both in the eprint and published version
of DIS3.

\section {Updates}
The energy \cite{DJS00,BF00,DJS01} and flux \cite {BFIJ01}
functions have now been computed up to order $v^7$
in post-Newtonian theory. The corresponding expansion coefficients
are as follows. The 3PN coefficients in the expansion of the
various energy functions are:
\begin{eqnarray}
\widehat E_3 & = & -\frac{675}{64} + \left [ \frac{34445}{576} 
           - \frac{205\pi^2}{96} + \frac{10\omega_s}{3} \right ]\eta \nonumber \\
       & - & \frac{155}{96}\eta^2 - \frac{35}{5184} \eta^3,
\end{eqnarray}
\begin{eqnarray}
e_3 & = &  -9 + \left ( \frac {4309}{72} - \frac {205}{96}\pi^2 
      +   \frac {10}{3}\omega_s \right )\eta \nonumber \\
    & - & \frac {103}{36}\eta^2 + \frac{1}{81} \eta^3,
\end{eqnarray}
with the Pad\'e approximant $e_{P_6}$ determined using Eq.~(2.17)
of DIS3, wherein $c_1$ and $c_2$ are as in DIS3 and $c_3$ is
given by
\begin{equation}
c_3= \frac {e_1e_3 - e_2^2}{e_1 \left ( e_1^2 - e_2 \right )}.
\end{equation}
The dimensionless parameter $\omega_s$ (used in \cite{DJS00}) is related to the 
parameter $\lambda$ (used in \cite{BF00}) by 
 $\omega_s =   -1987/840 - 11  \lambda/3$, so that we alternatively have
\begin{eqnarray}
\widehat E_3 & = & -\frac{675}{64} + \left [ \frac{209323}{4032} 
           - \frac{205\pi^2}{96} - \frac{110\lambda}{9} \right ]\eta \nonumber \\
       & - & \frac{155}{96}\eta^2 - \frac{35}{5184} \eta^3,
\end{eqnarray}
\begin{eqnarray}
e_3 & = &  -9 + \left ( \frac {26189}{504} - \frac {205}{96}\pi^2 
      -   \frac {110}{9}\lambda \right )\eta \nonumber \\
    & - & \frac {103}{36}\eta^2 + \frac{1}{81} \eta^3.
\end{eqnarray}
The numerical value of $\omega_s$ has been recently determined by 
dimensional regularization \cite{DJS01} to be simply equal to $\omega_s = 0 $,
 which corresponds to  $\lambda =   -1987/3080$.
[Note that there is a sign misprint in the second term on the right-hand-side of
Eq.(4.7) in the last reference in \cite{DJS00}; it should read
$\lambda = - 3 \omega_s/11 - 1987/3080$.]
 Concerning the 3PN update of the 
effective one-body Hamiltonian, it is explicitly given in 
section IVD of the second reference in \cite{DJS00}.

The expansion coefficients in Table II of DIS3 at 3PN and 3.5PN
are as follows. The coefficients in the expansion of the flux are:
\begin{eqnarray}
\widehat {\cal F}^v_{6} & = & \frac{6643739519}{69854400} + \frac{16 \pi^2}{3}
      - \frac {1712}{105} \gamma \nonumber \\ 
      & + & \left (- \frac {11497453}{272160} + \frac{41\pi^2}{48}
      + \frac{176}{9} \lambda - \frac{88}{3} \Theta \right ) \eta \nonumber \\
      & - &  \frac{94403}{3024} \eta^2 - \frac{775}{324} \eta^3,
\end{eqnarray}
\begin{equation}
\widehat {\cal F}^v_{7} =  \left (- \frac{16285}{504} + \frac{214745}{1728} \eta
               + \frac{193385}{3024} \eta^2 \right )\pi,
\end{equation}
and
$
\widehat {\cal F}^v_{l6} = -\frac{1712}{105}.
$
Here $\gamma$ is the Euler constant, $\gamma=0.577\ldots,$ and
$\Theta$ and $\lambda$ are two undetermined parameters in Ref. \cite{BFIJ01}. 
We use the letter $\Theta$ to denote what is denoted by $\theta$ in 
\cite{BFIJ01} ( this should not be confused with the related undetermined 3PN
quantity $\hat{\theta} = \theta -7 \lambda/3$ which is also used in some
formulas of \cite{BFIJ01}).
The $\lambda$ appearing in the flux function
 is the same quantity as in the energy function,
 arising,  as it does, from the  time derivatives
 of the mass quadrupole moment involved  in computing  the far-zone flux.
The quantity
$\widehat F^v_{l6}$ is the coefficient of the log term that arises, for
the first time, at the 3PN order; to the usual {\it Newton-normalized}
Taylor expansion \cite {DIS1} one must add 
$\widehat {\cal F}^v_{l6} \log(4 v) v^6$ to complete the PN expansion.
Beware that if expressions are rewritten in terms of $\omega_s$
rather than    $\lambda$,   the rational numerical coefficient in
the $\eta$ term will change. For ready reckoning, in the above
 and  subsequent formulas, we indicate this explicitly as follows:
($ -11497453/272160 + (176\lambda)/9 
\rightarrow -14930989/272160 - (16\omega_s)/3 $).  
 This means the  flux formula  may be alternatively
 written in terms of $\omega_s$ by the indicated replacement.

Coefficients in the expansion of time as a function of 
the invariant velocity parameter $v=(\pi m f)^{1/3},$ 
where $f$ is the gravitational-wave frequency, are
\begin{eqnarray}
 \hat t^v_6 
         & = & - \frac {10052469856691}{23471078400} 
           +   \frac {128}{3} \pi^2 \nonumber \\ 
         & + & \left ( \frac {15335597827}{15240960} - \frac {451}{12} \pi^2
           +   \frac {352}{3} \Theta - \frac {2464}{9}\lambda \right)  \eta \nonumber \\
         & + & \frac {6848}{105} \gamma - \frac {15211}{1728} \eta^2 
           +   \frac {25565}{1296} \eta^3,
\end{eqnarray}
\begin{equation}
 \hat t^v_7 =  \left(-\frac {15419335}{127008} - \frac {75703}{756}\eta 
           + \frac {14809}{378}  \eta^2\right)\pi, 
\end{equation}
and 
$
 \hat t^v_{l6} = \frac {6848}{105},
$
where $\hat t^v_{l6}$ is the coefficient of the log term;
to the usual Newton-normalized Taylor expansion one must add 
$\hat t^v_{l6} \log(4v) v^6$ to complete the PN expansion.
($15335597827/15240960 - (2464\lambda)/9
\rightarrow  18027490051/15240960 + (224\omega_s)/3$).

Coefficients in the expansion of the gravitational wave phase 
as a function of the invariant velocity $v$ are
\begin{eqnarray}
 \hat \phi^v_6 & = & \frac {12348611926451}{18776862720} - \frac {160}{3} \pi^2
           - \frac {1712}{21} \gamma \nonumber \\ 
           & + & \left (- \frac {15335597827}{12192768} + \frac {2255}{48} \pi^2
           + \frac {3080}{9} \lambda - \frac {440}{3} \Theta \right ) \eta \nonumber \\
           & + & \frac {76055}{6912} \eta^2 - \frac {127825}{5184} \eta^2,
\end{eqnarray}

\begin{equation}
 \hat \phi^v_7 = \left ( \frac {77096675}{2032128} + \frac {378515}{12096}\eta 
           - \frac {74045}{6048}\eta^2 \right ) \pi,
\end{equation}
and $ \hat \phi^v_{l6} = - \frac {1712}{21}, $
where, as in the previous cases, 
$\hat \phi^v_{l6}$ is the coefficient of the log term;
to the usual Newton-normalized Taylor expansion 
one must add $\hat \phi^v_{l6} \log(4v) v^6$ to 
complete the PN expansion.
($ -15335597827/12192768 + (3080\lambda)/9
\rightarrow
 -18027490051/12192768 - (280\omega_s)/3$).

The expansion coefficients of the phase as a function of 
the time parameter\footnote{
The dimensionless time variable $\theta$ here is related 
to $\tau$ in \cite{BF00}, by $\theta=\tau^{-1/8}$ leading to
minor  differences in the coefficients here and in \cite{BF00}. }
 $\theta =[\eta (t_{\rm ref} - t )/(5m)]^{-1/8}$ 
where $t_{\rm ref}$ is a reference time at which the PN-expanded GW frequency 
 formally goes to infinity,
are given by,
\begin{eqnarray}
 \hat \phi^t_6 
         & =  & \frac {831032450749357}{57682522275840} - \frac {53}{40}\pi^2
           - \frac {107}{56} \gamma \nonumber \\ 
         & +  & \left (- \frac {123292747421}{4161798144}
           + \frac {2255}{2048} \pi^2 + \frac {385}{48} \lambda
           - \frac {55}{16} \Theta \right ) \eta \nonumber \\
         & + & \frac {154565}{1835008} \eta^2
           - \frac {1179625}{1769472} \eta^3.
\end{eqnarray}
\begin{equation}
 \hat \phi^t_7 =  \left ( \frac {188516689}{173408256} 
               + \frac {488825}{516096} \eta
               - \frac {141769}{516096} \eta^2 \right )\pi,
\end{equation}
and $ \hat \phi^t_{l6} = -\frac {107}{56} $,
where, as before, $\hat \phi^t_{l6}$ is the coefficient of the log term;
to the usual Newton-normalized Taylor expansion one must add 
$\hat \phi^t_{l6} \log(2\theta) \theta^6$ to complete the PN expansion.
($ -123292747421/4161798144 + (385\lambda)/48
\rightarrow
 -144827885213/4161798144 - (35\omega_s)/16$).

Finally, coefficients in the expansion of the gravitational-wave frequency 
in terms of the time parameter $\theta$ are given by
\begin{eqnarray}
\widehat F^t_{6} & = & - \frac {720817631400877}{288412611379200} 
           + \frac {53}{200}\pi^2
           + \frac {107}{280} \gamma  \nonumber \\ 
           & + & \left ( \frac {123292747421}{20808990720} - \frac{451}{2048} \pi^2
           - \frac {77}{48} \lambda + \frac {11}{16} \Theta \right ) \eta \nonumber \\
           & - & \frac {30913}{1835008} \eta^2
           + \frac {235925}{1769472} \eta^3, 
\end{eqnarray}
\begin{equation}
\widehat F^t_{7} = \left (- \frac {188516689}{433520640} - \frac{97765}{258048} \eta
           + \frac {141769}{1290240} \eta^2 \right ) \pi.
\end{equation}
and
$
\widehat F^t_{l6} = \frac {107}{280},
$
where, $\widehat F^t_{l6}$ is the coefficient of 
the log term;
to the usual Newton-normalized Taylor expansion 
one must add $\widehat F^t_{l6} \log(2\theta) \theta^6$ 
to complete the PN expansion.
($ 123292747421/20808990720 - (77\lambda)/48
\rightarrow
 144827885213/20808990720 + (7\omega_s)/16$).

In computing Pade coefficients of the {\it new} flux function \cite {DIS1}
one needs the first
seven continued fraction coefficients. The first six of these
are as in Appendix~A of Ref.~\cite{DIS1}, except that the last
term in the first line of $c_6$ should be $c_3^2(c_2-c_1).$ The
coefficient $c_7$ is too long to be quoted in this brief note; an 
electronic version can be obtained from the authors.
 
\acknowledgements

We would like to thank J-Y. Vinet for pointing out the missing factor in Eq.~(4.5) of
DIS3 and L. Blanchet for help received in confirming 3PN and 3.5PN expansion
coefficients.
BSS would like to thank Max-Planck Institute for Gravitational Physics,
 Albert Einstein Institute, for hospitality, where this Brief Communication
 was written.

\end {document}